\documentclass[twocolumn]{aastex631}

\usepackage{color,comment,multirow,booktabs,upgreek}
\usepackage[hang,flushmargin]{footmisc} 
\usepackage{amsmath}

\graphicspath{{./}{images/}}

\newcommand{\hi}{\ifmmode{\rm HI}\else{H\/{\sc i}}\fi} 
\newcommand{\vlos}{\ifmmode{V_\mathrm{los}}\else{$V_\mathrm{los}$}\fi}
\newcommand{\vsys}{\ifmmode{V_\mathrm{sys}}\else{$V_\mathrm{sys}$}\fi}
\newcommand{\vrot}{\ifmmode{V_\mathrm{rot}}\else{$V_\mathrm{rot}$}\fi}
\newcommand{\vrad}{\ifmmode{V_\mathrm{rad}}\else{$V_\mathrm{rad}$}\fi}
\newcommand{\vdisp}{\ifmmode{\sigma_\mathrm{gas}}\else{$\sigma_\mathrm{gas}$}\fi}
\newcommand{\vflat}{\ifmmode{V_\mathrm{flat}}\else{$V_\mathrm{flat}$}\fi} 
\newcommand{\de}{\ifmmode{^\circ}\else{$^\circ$}\fi} 
\newcommand {\kms}{\ifmmode{\rm km \, s^{-1}}\else{$\rm km \, s^{-1}$}\fi}

\newcommand*{\code}[1]{\ensuremath{\mathtt{#1}}}

\renewcommand{\eqref}[1]{Eq.\ (\ref{#1})}

\newcommand{\mstar}{\ifmmode{M_{\star}}\else{$M_{\star}$}\fi}

\shorttitle{Shreds in galaxy catalogs}
\shortauthors{Di Teodoro et al.}

\begin{document}

\title{\large Identification of galaxy shreds in large photometric catalogs \\ using Convolutional Neural Networks}%

\correspondingauthor{E.~M. Di Teodoro}
\email{enrico.diteodoro@unifi.it}

\author[0000-0003-4019-0673]{Enrico M.\ Di Teodoro}
\affiliation{Department of Physics \& Astronomy, Johns Hopkins University, Baltimore, MD 21218, USA}
\affiliation{Space Telescope Science Institute, 3700 San Martin Drive, Baltimore, MD 21218, USA}
\affiliation{Dipartimento di Fisica e Astronomia, Università degli Studi di Firenze, 50019 Sesto Fiorentino, Italy}

\author[0000-0003-4797-7030]{J.\ E.\ G.\ Peek}
\affiliation{Space Telescope Science Institute, 3700 San Martin Drive, Baltimore, MD 21218, USA}
\affiliation{Department of Physics \& Astronomy, Johns Hopkins University, Baltimore, MD 21218, USA}

\author[0000-0002-5077-881X]{John F.\ Wu}
\affiliation{Space Telescope Science Institute, 3700 San Martin Drive, Baltimore, MD 21218, USA}
\affiliation{Department of Physics \& Astronomy, Johns Hopkins University, Baltimore, MD 21218, USA}

\begin{abstract}
Contamination from galaxy fragments, identified as sources, is a major issue in large photometric galaxy catalogs. 
In this paper, we prove that this problem can be easily addressed with computer vision techniques. 
We use image cutouts to train a convolutional neural network (CNN) to identify catalogued sources that are in reality just star formation regions and/or shreds of larger galaxies. 
The CNN reaches an accuracy $\sim 98\%$ on our testing datasets.
We apply this CNN to galaxy catalogs from three amongst the largest surveys available today: the Sloan Digital Sky Survey (SDSS), the DESI Legacy Imaging Surveys and the Panoramic Survey Telescope and Rapid Response System Survey (Pan-STARSS).
We find that, even when strict selection criteria are used, all catalogs still show a $\sim5\%$ level of contamination from galaxy shreds.
Our CNN gives a simple yet effective solution to clean galaxy catalogs from these contaminants. 
\end{abstract}


\section{Introduction}
\label{sec:intro}
In the past thirty years, large blind surveys of the sky with modern telescopes have revolutionized our understanding of galaxy formation and evolution.
From early optical surveys, like the Digitized Sky Survey (DSS), to the latest surveys, like the Sloan Digital Sky Survey \citep[SDSS,][]{Eisenstein+2011}, the Panoramic Survey Telescope and Rapid Response System Survey  \citep[Pan-STARRS,][]{Chambers+2016} or the Dark Energy Spectroscopic Instrument (DESI) Legacy Imaging Surveys \citep[][]{Dey+2019}, a huge number of astronomical images in multiple photometric bands and in multiple epochs have been produced throughout the years and have allowed astronomers worldwide to study the properties of galaxies on large statistical samples.

A significant fraction of the science done with these surveys is carried out using large catalogs of objects extracted from the astronomical images, rather than with the images themselves.
For this reason, over the years, a lot of effort has been put into developing reliable, efficient and fully-automated source finding algorithms, able to identify robustly sources and estimating their most important photometric parameters, like magnitudes, colors, redshifts and sizes.
Well-tested codes, like the Source Extractor \citep[SExtractor,][]{Bertin+1996}, the Tractor \citep{Lang+2016} and ProFound \citep{Robotham+2018}, are routinely used to extract source catalogs from astronomical images \citep[see][for a review of different source finding techniques tested in the years]{Masias+2012,Masias+2013}.

\begin{figure*}
    \center
    \includegraphics[width=\textwidth]{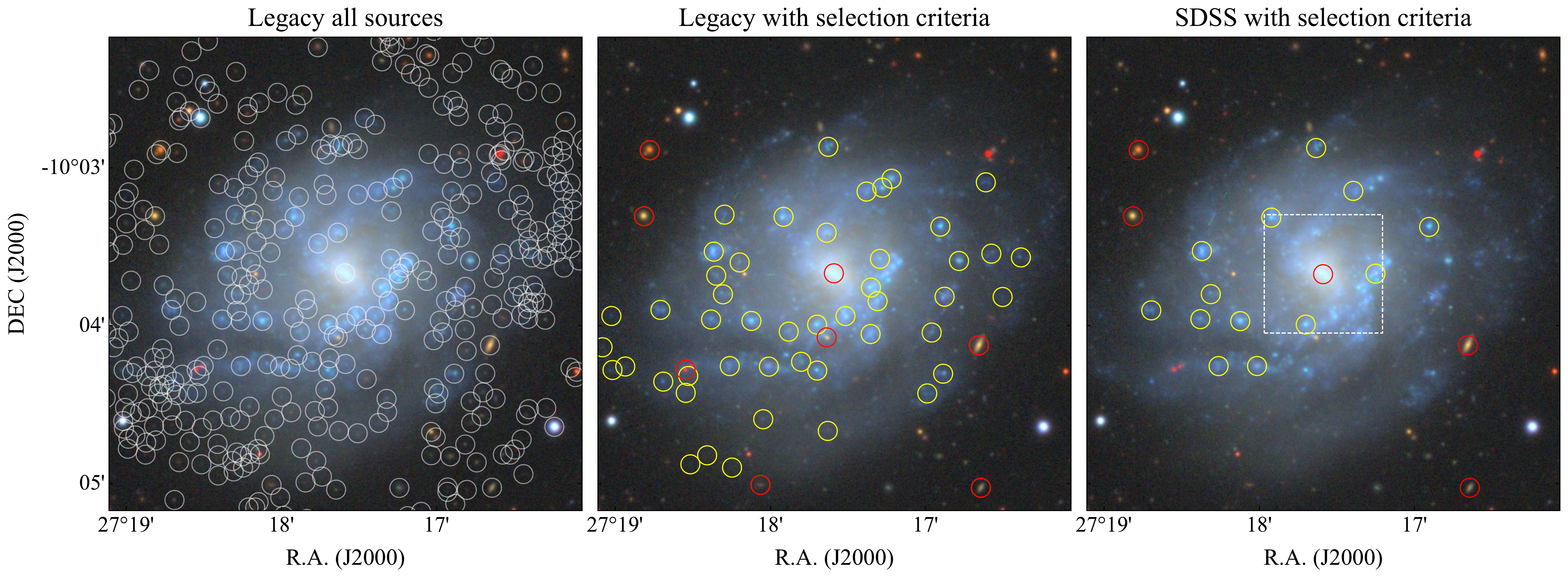}
    \caption{Photometric sources around the spiral galaxy UGCA 021. In all panels, we show a color image in $g$-$r$-$z$ band from the Legacy surveys. In the left panel, we show all sources in the Legacy DR9 catalog, the middle and right panels show sources in Legacy and SDSS DR17 catalogs, respectively, after some cleaning using stricter selection criteria, similar to those described in Section~\ref{sec:sample}. In the middle and right panels, galaxy shreds are highlighted in yellow, while genuine galaxy sources in red. The only NSA spectroscopically-confirmed source in this field is the galaxy UGCA 021. Images have $512\times512$ pixels with a pixel size of $0.262''$. The white-dashed square in the right panel is an example of a $128\times128$ pixel image used by our CNN.} 
    \label{fig:example}
\end{figure*} 

One of the problems that still affects modern photometric catalogs is the considerable presence of galaxy shreds, i.e.\ spatially extended galaxies are often broken down into multiple objects. 
It is clear that this issue is more prominent in relatively low-redshift galaxies, which can cover larger angular regions and can show several resolved star-formation regions in their disks. 
An example of this is illustrated in \autoref{fig:example} for the nearby galaxy UGCA 021. 
Leftmost panel shows all sources (white circles) in the field that are listed in the photometric catalog of the Legacy survey, based on source extraction with the Tractor. 
The vast majority of sources in the catalog are in reality star-formation regions within the disk of the galaxy.
Catalogs can be partially cleaned by using more aggressive selection criteria, for example based on colors and sizes of detected sources.
The central and right panels of \autoref{fig:example} show objects left after more stringent criteria are applied to Legacy and SDSS catalogs, respectively (see Section~\ref{sec:sample}). 
Although many spurious detections have been removed, there are still more shreds of the galaxy (yellow circles) than genuine sources (red circles). 
Shreds can be more or less numerous in different photometric catalogs, as shown in the left panel for the SDSS catalog.

In this paper, we present and make available a new method to clean catalogs from galaxy shreds. 
A convolutional neural network (CNN) is trained on three-band optical images in order to identify whether objects in catalogs are galaxy fragments or genuine sources. 
The remainder of this paper is structured as follows. 
Section~\ref{sec:methods} describes the data used in this work and the machine-learning algorithm used for identifying galaxy shreds. 
We apply our CNN to some of the most complete photometric catalogs available nowadays and discuss our findings in Section~\ref{sec:results}. 
We summarize and conclude in Section~\ref{sec:conclusions}.
The code used in our analysis and the trained CNN model are publicly available on Github: \href{https://github.com/editeodoro/CNN_shreds}{https://github.com/editeodoro/CNN\_shreds}. \\\\

\section{Methods}
\label{sec:methods}

\subsection{Data}
We used three-band images from the Legacy Surveys to train and test our machine learning algorithm.
The DESI Legacy Imaging Surveys \citep[or Legacy, for simplicity,][]{Dey+2019} is a combination of three public projects with DESI, i.e.\ the DECam Legacy Survey (DECaLS), the Beijing-Arizona Sky Survey \citep[BASS,][]{Zou+2017} and Mayall $z$-band Legacy Survey (MzLS).
Legacy provides imaging of 14,000 deg$^2$ of the northern sky in three optical or near-infrared filters, i.e.\ $g$-$r$-$z$ bands, covering most of the SDSS footprint with a comparable spatial resolution but significantly deeper integration times than SDSS. 

\begin{figure*}
    \center
    \includegraphics[width=\textwidth]{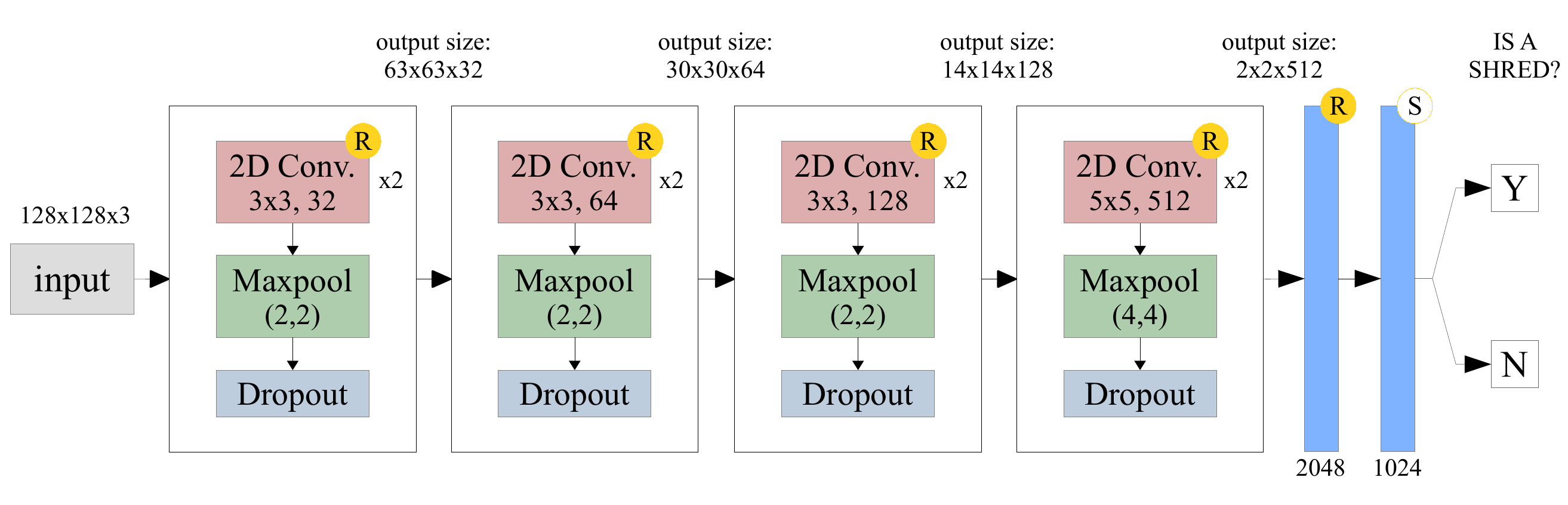}
    \caption{A schematic of the architecture of the convolutional neural network used in this work. Our CNN has 10 layers, including 8 2D convolutional layers (red boxes) and 2 fully-connected layers (blue tall boxes), with approximately 10.6M parameters. 
    Max pooling and dropout operations are denoted as green and blue boxes, respectively.
    In convolutional layers, we indicate the size of the 2D windows and the number of output filters.
    In pooling stages, we also write the size of downsampling.
    Activation functions are indicated in the circles (R for ReLU and S for softmax function).\\}
    \label{fig:cnn}
\end{figure*}

We put together a dataset for training/testing by starting from the full Legacy DR9 photometric catalog\footnote{Available at \href{https://datalab.noirlab.edu/query.php}{datalab.noirlab.edu/query.php}.}. 
Because galaxy shreds will be dominant in low-redshift systems, we used the NASA-Sloan Atlas (NSA) to reduce the pool of candidate systems. 
The NSA is a catalog of spectroscopically confirmed galaxies at redshift $z<0.15$ \citep{Blanton+2011}.
In particular, we used v1.0.1 of the NSA catalog\footnote{Available at \href{https://data.sdss.org/sas/dr17/sdss/atlas/v1}{data.sdss.org/sas/dr17/sdss/atlas/v1}.}.
We only kept objects in the Legacy catalog that are within 100 kpc from a NSA source, leaving us with $\sim$2,000,000 possible targets.  
Amongst these, we created a training set of 5,000 sources by visually inspecting their images and by manually labelling them as either a galaxy or a shred. 
The visual classification of sources took into account a number of features that allow a human brain to distinguish between a genuine galaxy and a shred; this includes, for example, the color, shape, size and coherence of a source with respect to the surrounding region of an image.
We selected sources such that, in the final training set, objects are split approximately $60\%-40\%$ between real galaxies and galaxy shreds.

To generate the images used by the deep learning algorithms, we downloaded image cutouts of our training objects from the Legacy survey viewer website\footnote{\href{https://www.legacysurvey.org/viewer}{www.legacysurvey.org/viewer}}.
We fetched images in the $g$-$r$-$z$ filters in the Portable Network Graphics (PNG) format at a 0.262 pix arcsec$^{-1}$ resolution.
In particular, we used $128\times128$ pixel images, corresponding to about $34'' \times 34''$ per side.
Image cutouts were centered on the position centroid of each object in our training set.

\subsection{Convolutional Neural Networks}
We adopted a machine-learning algorithm to identify galaxy shreds from multi-band images. 
Individual star-formation regions within a galactic disk have very distinctive morphological features with respect to a galaxy as a whole. 
Convolutional Neural Networks \citep{Lecun+98,Lecun+2015} are ideal to identify these features and classify objects
based on their appearance.
A CNN is a particular type of neural network that includes a number of layers performing convolution, which can be used to extract weighted features from input images for a given problem.
In particular, our task is a simple binary classification problem, i.e.\ whether a target is either a proper galaxy or just a shred of a galaxy.

The architecture of the CNN used in this paper is illustrated in \autoref{fig:cnn}.
Our design is inspired by the VGG19 architecture \citep{Simonyan+2014}, a commonly used CNN for large-scale image recognition, but with fewer convolutional layers and parameters. 
Our network includes 8 convolutional layers (red boxes) and 2 fully-connected layers (blue tall boxes), for a total of 10 layers and approximately 10.6 million trainable parameters.
The dimension of inputs is $128\times128\times3$, with the depth denoting the three image bands ($g$-$r$-$z$).
Images go through four main stages of convolution with kernel sizes 3, 3, 3 and 5, and 32, 64, 128 and 512 filters, respectively. 
Multiple convolution layers with small convolution kernels increase the effective receptive field and add more representational flexibility to the model.
A max-pooling layer (green boxes) follows convolution to downsample the matrices and to reduce the spatial size and the number of parameters. 
Sizes of max-pooling are 2, 2, 2 and 4, respectively. 
Dropouts ($\rho=0.2$, blue boxes) are applied in each stage to discard unnecessary parameters and to prevent overfitting during the training of the CNN. 
Finally, the two fully-connected dense layers with 2048 and 1024 hidden units produce the binary classification into galaxy or shred.

For all convolutional layers and the first dense layer, we used a Rectified Linear Unit \citep[ReLU,][]{Nair2010} activation function, i.e.\ $f(x)=\mathrm{max(0,x)}$.
The last output layer uses instead a ``softmax'' function \citep{Bishop2006}, $f(x) = \exp(x)/\sum_j \exp(x_j)$, which provides a sort of probability for each class $j$.
Therefore, for each target, our CNN predicts an estimate ($p_\mathrm{CNN}$) of how likely the object is to be a shred. 
For our binary problem, we classify objects with $p_\mathrm{CNN}\geq0.5$ as shreds and objects with $p_\mathrm{CNN}<0.5$ as galaxies.
We optimize the CNN by minimizing the binary cross entropy loss.

We trained the CNN illustrated in \autoref{fig:cnn} using the dataset of 5,000 objects described in previous Section.
During the training phase, we used a RMSprop optimizer with plain momentum and a learning rate of 0.001.
A $k$-fold cross-validation with $k=5$ was used to better evaluate how well the CNN is performing on independent datasets.
We first reserved 20\% of the data for testing. 
The remaining 80\% was split into five random subsets, each including a training set (80\%) and a validation set (20\%), which was used to benchmark the predictions of the CNN. 
Moreover, we used image augmentation to artificially increase the number of data in the training set: new input images were created by randomly rotating and/or flipping the original dataset, which also helps the CNN to learn translational and rotational symmetry \citep[][]{Dieleman+2015}.
During training, we set a maximum number of 100 epochs, with an early stopping mechanism based on the trend of the loss function of the validation set, i.e.\ training stops when the validation loss function does not decrease over 5 consecutive epochs. 
The CNN typically achieves convergence after 20-25 epochs. 

\begin{figure}
    \centering
    \includegraphics[width=0.48\textwidth]{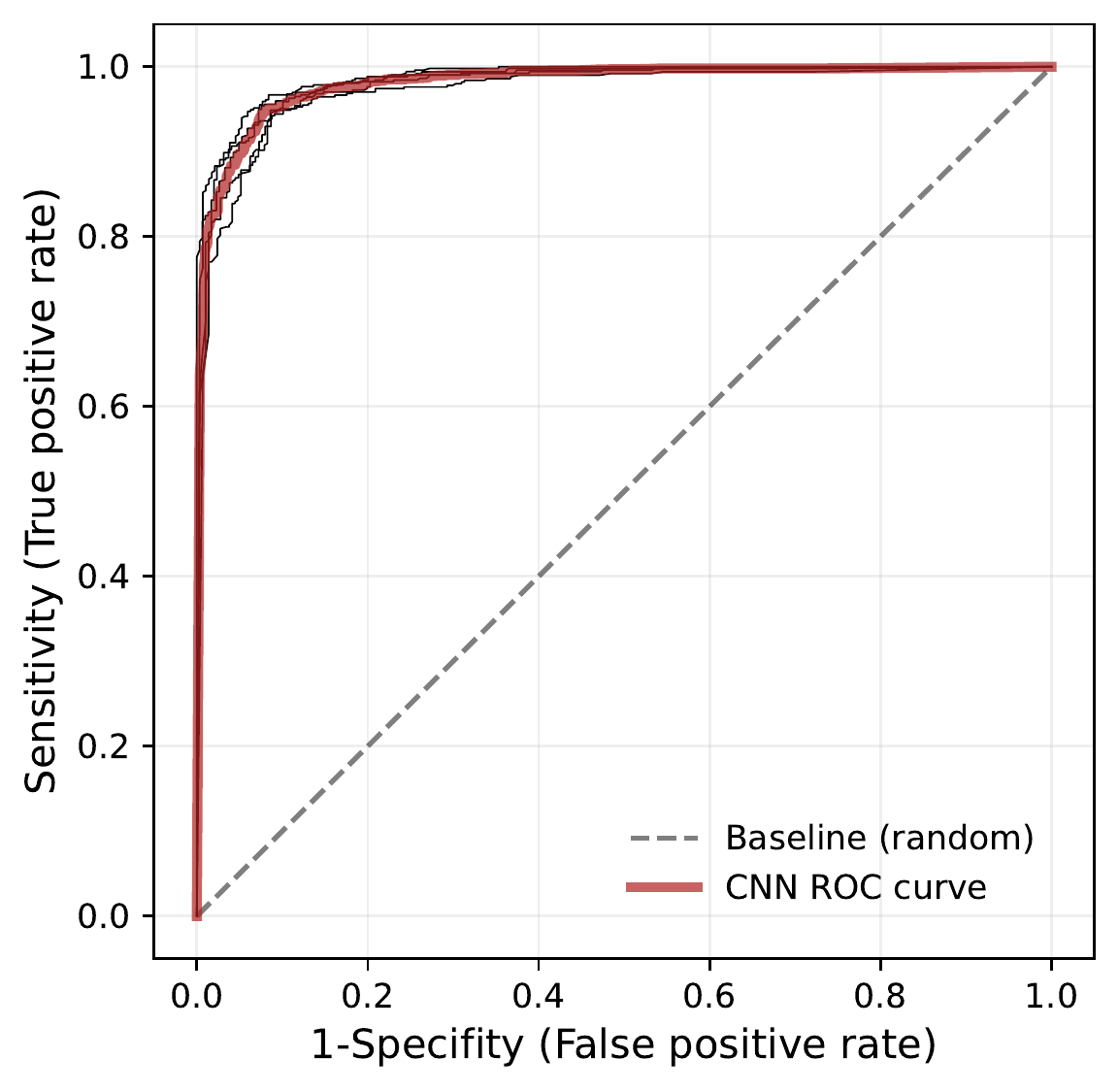}
    \caption{The receiving operator characteristic (ROC) curve for our CNN model. Black thin lines denote the five cross-validation folds used, while the mean result is shown as a thick red line. The AUC for our model is 0.985. The dashed grey line shows the baseline for a random guess (AUC = 0.5).}
    \label{fig:roc}
\end{figure}

\autoref{fig:roc} shows the receiving operator characteristic (ROC) plot for our model, i.e.\ a curve of the true positive rate against the false positive rate for various CNN probability thresholds $p_\mathrm{CNN}$. The true positive rate, also referred to as sensitivity or completeness, is $\mathcal{C} \equiv \mathrm{TP/(TP + FN)}$ where TP is the true number of shreds and FN is the number of missed shreds.
The false positive rate is $(1-\mathcal{S})$, related to the specificity $\mathcal{S} = \mathrm{TN / (TN + FP)}$, where TN is the true number of galaxies and FP is the number of missed galaxies.
The ROC can be used to evaluate the performance of a model: a good model will have a large area under the ROC curve (AUC), i.e.\ it will be able to maximize the true positive rate and, at the same time, to minimize the false positive rate. 
A perfect, ideal model would have an AUC of 1, while random guesses (grey dashed line in \autoref{fig:roc}) have $\mathrm{AUC}=0.5$. 
Our CNN model, shown as a red thick line, has an $\mathrm{AUC}=0.985$, indicating that our CNN has a high efficiency in discriminating between real galaxy and galaxy shreds. 
The accuracy $\mathcal{A} = \mathrm{(TP + TN) / (TP + TN + FP + FN)}$ of our CNN model is $0.97-0.98$.\\

\section{Applications}
\label{sec:results}

In this section, we use our trained CNN to identify galaxy shreds in photometric catalogs from three well-known surveys: SDSS DR17\footnote{Available at \href{http://skyserver.sdss.org/dr17/}{skyserver.sdss.org/dr17}.}, Legacy DR9 and Pan-STARSS PS1 DR2\footnote{Available at \href{https://mastweb.stsci.edu/ps1casjobs}{mastweb.stsci.edu/ps1casjobs}.}. 
We build galaxy catalogs from each survey using appropriate selection criteria and we investigate the level of contamination from galaxy shreds. 

\subsection{Example catalog selection}
\label{sec:sample}
Photometric catalogs created with automated software usually includes sky objects that can be either galaxies or stars. 
In addition to these real objects, they can include fragments of galaxies and other bad sources, for example image artifacts that are mistakenly catalogued as sources. 
Depending on the science goals, a list of galaxy candidates can be usually extracted from the entire galaxy catalog by imposing appropriate photometric and quality cuts. 
To test our CNN, we built a source sample by querying photometric catalogs from SDSS, Legacy and Pan-STARSS surveys with a number of standard parameters.

As an example of a possible scientific case, we used selection criteria inspired by the cuts applied to select host galaxy candidates for the Satellites Around Galactic Analogs survey \citep[SAGA,][]{Geha+2017}.
SAGA is searching for dwarf galaxies around Milky-Way-like hosts, thus it is especially susceptible to contamination by host shredding.
We note that this is just a pedagogical example to show how our CNN could be usefully applied to real case studies. 
Our main sample selection was done by means of simple cuts in the surface brightness–magnitude and color–magnitude planes \citep{Mao+2021,Wu+2022}:
\begin{align*}
r &\leq 21.0, \\
\mu_\mathrm{eff} + \sigma_{\mu} - 0.7 \, (r - 14) &> 18.5, \\
(g-r) - \sigma_{gr} + 0.06\,(r - 14) &< 0.90, \\
\end{align*}
\noindent where $\mu_\mathrm{eff}\equiv r +2.5\log(2\pi R_{r,\mathrm{eff}})$ is the effective surface brightness in $r$-band, calculated from the extinction-corrected magnitude $r$ and the half-light radius $R_{r,\mathrm{eff}}$. In the above cuts, $(g-r)$ is the extinction-corrected color, while $\sigma_\mu$ and $\sigma_{gr}$ are errors on the effective surface brightness and on the color, respectively.

Beside these magnitude, color, and surface brightness cuts, we also applied a number of selection and quality flags to start cleaning our catalogs. 
Here, we only describe parameters for the Legacy catalog, but analogous cuts were applied to all three surveys.
For the Legacy survey, first of all we used the morphological flag \code{TYPE\neq PSF} and \code{TYPE\neq DUB} to select only galaxies and to reject stars\footnote{See \href{https://www.legacysurvey.org/dr9/catalogs/}{www.legacysurvey.org/dr9/catalogs} for details on the catalog quantities.}. 
For the same reason, we also required that objects have measured half-light radius in $r$ band (\code{SHAPE\_R>0}).
A series of quality flags was then applied to define a sample of ``good'' galaxy targets, i.e.\ objects that can be well described in terms of a simple galaxy model (exponential, deVacouleur or Sérsic profiles):
\begin{align*}
&\code{NOBS} \geq 1, \\
&\code{DERED\_MAG \neq nan}, \\
&\code{ALLMASK} = 0, \\
&\code{FRACMASKED} < 0.35, \\
&\code{FRACFLUX} < 4, \\
&\code{RCHISQ} < 10, \\
&\code{RCHISQ} < 4 \text{ (any one band)},\\
&\code{FRACIN} > 0.7 \text{ (any one band), and}\\
&\sigma(\text{magnitude}) < 0.2.
\end{align*}

\noindent where the above criteria are applied to all three bands ($g$-$r$-$z$), unless otherwise noted.
The first two criteria require the presence of good measurements in all bands, the third criterion is a standard quality mask for Legacy catalogs. 
The other criteria reject sources that are not well described by the model due, for example, to bad fits and/or considerable source blending. 
We note that these criteria for selecting ``good'' galaxy targets can be considered quite aggressive and should already remove many spurious detections.

Because we know that galaxy shreds will be found within a certain distance from the center of relatively low-redshift galaxies, in each catalog we only kept sources within a projected distance of 100 kpc from a spectroscopically-confirmed galaxy in the NSA catalog, similarly to what we did for the training dataset.
Finally, because our CNN is trained with $g$-$r$-$z$ band images from the Legacy surveys, we discard all sources that have either corrupted images or that are not covered in the Legacy footprint (for the SDSS and Pan-STARSS catalogs).
This leaves us with three catalogs containing approximately 800,000 (SDSS and Pan-STARSS) or 700,000 (Legacy) galaxy candidates.

\begin{figure*}
    \center
    \includegraphics[width=\textwidth]{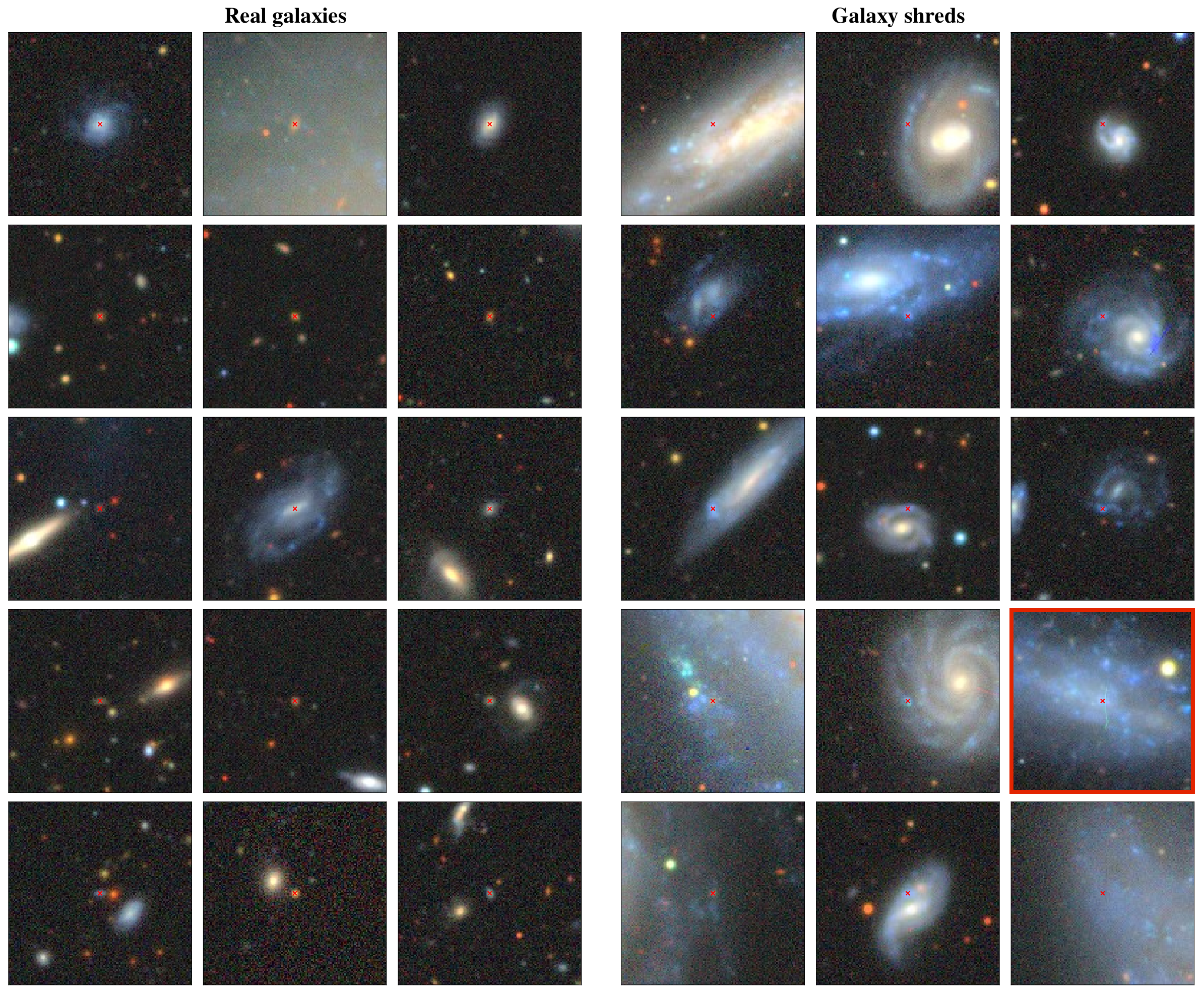}
    \caption{Binary classification of sources into either a galaxy or a shred using the CNN of \autoref{fig:cnn}. The left group of panels shows a random sample of targets identified as genuine galaxies, the right group of panels a random sample of targets classified as galaxy shreds. The source in a red frame highlights a (possibly) wrong classification, i.e.\ a galaxy likely mislabelled as a shred by the CNN.}
    \label{fig:results}
\end{figure*}

\subsection{Shreds in catalogs}
We downloaded $128\times128$ color images from the Legacy survey centered on each source in the three catalogs. 
We fed our trained CNN with these images and obtained a binary classification into galaxy-shred for each object. 
\autoref{fig:results} illustrates 15 randomly-selected examples of objects in each class from the Legacy catalog: the left group of panels shows sources classified as genuine galaxies, while objects rather identified as galaxy shreds are shown in the right panels. 
Red crosses in all images indicate the centroid position of each source, according to our galaxy catalogs. 
Images in \autoref{fig:results} confirm that our CNN is extremely efficient and powerful in this particular classification problem, being able to disentangle real galaxies from shred in most cases. 

The only questionable choice, from a human eye perspective, is the source highlighted with a red frame amongst the galaxy shreds. 
This is a flocculent star-forming 
galaxy that does not have a concentration of light in the central regions, which likely led the CNN to classify it as a simple shred.  
All other objects identified as shreds are, as expected, star-forming regions within larger discs that are misclassified as galaxies in the photometric catalogs. 
We note that our CNN in some cases is also able to deblend galaxies in background that overlap with much closer galaxies in foreground.
An example of this can be seen in the first row of images amongst the real galaxy column in \autoref{fig:results}.
However, we stress that our CNN is not purposefully trained for this and that several more advanced algorithms for galaxy deblending have been developed in the latest years \citep[e.g.,][]{Reiman+2019,Arcelin+2021,Hausen+2022}. 
As a matter of fact, from a visual inspection of several hundred sources classified by our CNN, we realized that most of misclassified objects are actually compact high-redshift galaxies overlapping with low-redshift galaxies that our CNN labels as shreds.

In summary, we found that shreds make up for approximately 5\% of sources in the SDSS and Pan-STARSS catalogs, and 4\% of sources in the Legacy catalog.
Therefore, despite the strict SAGA-like criteria applied to build these catalogs, we still observed a non-negligible contamination from galaxy shreds. 
We stress that the percentage of these contaminants can significantly increase if more relaxed criteria are used to build galaxy catalogs. 
For instance, simply removing the quality cut on \code{RCHISQ} for the Legacy photometric selection (see Section~\ref{sec:sample}) makes the fraction of shreds to rise from $\sim4\%$ to $\sim7\%$. \\

\section{Summary and conclusions}
\label{sec:conclusions}
In this work, we proposed a simple solution to a well-known problem that affects all photometric catalogs, i.e.\ the fact that extended galaxies are often shredded into multiple objects.
The ability of quickly recognizing galaxy shreds is fundamental to clean up large galaxy catalogs with million sources. 
To this end, we trained a 10-layer convolutional neural network (CNN) to classify objects into either genuine galaxies or shreds, starting from three-band ($g$-$r$-$z$) color images from the Legacy surveys. 
The CNN was able to identify reliably galaxy shreds, reaching an accuracy of $\sim98\%$ on the testing dataset.
Our trained CNN model is made available to the community at \href{https://github.com/editeodoro/CNN_shreds}{https://github.com/editeodoro/CNN\_shreds}. 
We stress that our CNN can be easily modified to be trained and to work with images from any other optical/infrared survey, for example with five-band images from SDSS or Pan-STARSS surveys.

Such a CNN model can be useful for several scientific applications.
We exemplify the potentiality of this approach by applying our CNN to galaxy catalogs built with selection criteria analogous to those used for choosing targets for the recent SAGA survey. 
These criteria are particularly aggressive and should in theory already dismiss many contaminants.
In particular, we built three galaxy catalogs, each containing some $\sim800$K objects, starting from general photometric catalogs of the Legacy, SDSS and Pan-STARSS surveys. 
We used then the trained CNN to classify these photometric sources based on their color images. 
Our CNN returned a fraction of shreds of $\simeq 5\%$ in each catalogs, highlighting how a relatively large number of spurious detections still affects these galaxy catalogs.
In conclusion, our work demonstrates that CNNs are a powerful and efficient tool to identify contaminants and remove them easily from galaxy catalogs.

\vspace*{0.5cm}
\section*{Acknowledgments}
EDT was supported by the US National Science Foundation under grant 1616177 and by the European Research Council (ERC) under grant
agreement no. 101040751.
This work made use of data from the Legacy Surveys, from the Sloan Digital Sky Survey IV (SDSS) and from the Panoramic Survey Telescope and Rapid Response System Survey (Pan-STARSS).
The Legacy Surveys consist of three individual and complementary projects: the Dark Energy Camera Legacy Survey (DECaLS; Proposal ID 2014B-0404), the Beijing-Arizona Sky Survey (BASS; NOAO ID 2015A-0801), and the Mayall z-band Legacy Survey (MzLS; ID 2016A-0453).
Funding for the SDSS IV has been provided by the Alfred P.\ Sloan Foundation, the U.S.\ Department of Energy Office of Science, and the Participating Institutions. 
SDSS-IV acknowledges support and resources from the Center for High Performance Computing  at the University of Utah. 
SDSS-IV is managed by the Astrophysical Research Consortium 
for the Participating Institutions of the SDSS Collaboration.
The Pan-STARRS1 Surveys (PS1) and the PS1 public science archive have been made possible through contributions by the Institute for Astronomy, the University of Hawaii, the Pan-STARRS Project Office, the Max-Planck Society and its participating institutes, the Max Planck Institute for Astronomy, Heidelberg and the Max Planck Institute for Extraterrestrial Physics, Garching, The Johns Hopkins University, Durham University, the University of Edinburgh, the Queen's University Belfast, the Harvard-Smithsonian Center for Astrophysics, the Las Cumbres Observatory Global Telescope Network Incorporated, the National Central University of Taiwan, the Space Telescope Science Institute, the National Aeronautics and Space Administration under Grant No. NNX08AR22G issued through the Planetary Science Division of the NASA Science Mission Directorate, the National Science Foundation Grant No. AST-1238877, the University of Maryland, Eotvos Lorand University (ELTE), the Los Alamos National Laboratory, and the Gordon and Betty Moore Foundation.


\software{AstroPy \citep{astropy:2013,astropy:2018}, matplotlib \citep{Hunter+2007}, TensorFlow \citep{tensorflow2015-whitepaper}, Keras \citep{chollet2015}. }

\bibliography{biblio}{}
\bibliographystyle{aasjournal}


\end{document}